\documentclass[reprint, aps, prx, longbibliography, twocolumn]{revtex4-2} 
\usepackage[utf8]{inputenc}
\usepackage{braket}
\usepackage{amsmath}
\usepackage{amssymb}
\usepackage{float}
\usepackage{graphicx}
\usepackage[format=plain]{subcaption}
\usepackage{xcolor}
\usepackage{url}
\usepackage{bm}
\usepackage{tikz}
\usepackage{tcolorbox}
\usetikzlibrary{shapes.multipart}
\usetikzlibrary{arrows.meta}
\usetikzlibrary{calc}
\newcommand{\mb}[1]{\mathbf{#1}}
\newcommand{\ul}[1]{\underline{#1}}
\newcommand{\beq}{\begin{equation}}
\newcommand{\eeq}{\end{equation}}

\newcommand{\RR}{\mathbb{R}}
\newcommand{\sigmab}{\ul{\sigma}}
\newcommand{\xb}{\ul{x}}
\newcommand{\Ab}{\mathbf{A}}
\newcommand{\s}{\mathcal{S}}
\newcommand{\thetab}{\ul{\theta}}

\long\def\ca#1\cb{} 

\makeatletter
\newcommand*{\balancecolsandclearpage}{%
  \close@column@grid
  \clearpage
  \twocolumngrid
}
\makeatother

\renewcommand{\figurename}{Fig.}

\begin{document}

\title{Learning Energy Based Representations of Quantum Many-Body States}

\author{Abhijith Jayakumar$^{1,2}$, Marc Vuffray$^{1}$, Andrey Y. Lokhov$^{1}$}
\affiliation{$^{1}$Theoretical Division, Los Alamos National Laboratory, Los Alamos, NM 87545, USA}
\affiliation{$^{2}$Center for Nonlinear Studies, Los Alamos National Laboratory, Los Alamos, NM 87545, USA}

\begin{abstract}
    Efficient representation of quantum many-body states on classical computers is a problem of enormous practical interest. An ideal representation of a quantum state combines a succinct characterization informed by the system's structure and symmetries, along with the ability to predict the physical observables of interest. A number of machine learning approaches have been recently used to construct such classical representations \cite{carleo2017solving, torlai2018latent, torlai2019wavefunction, melkani2020eigenstate, vicentini2022positive, schmale2022efficient}  which enable predictions of observables \cite{carrasquilla2019reconstructing} and accounts for physical symmetries \cite{morawetz2021u}. However, the structure of a quantum state gets typically lost unless a specialized ansatz is employed based on prior knowledge of the system \cite{cramer2010efficient, gross2010quantum, lanyon2017efficient, rocchetto2018stabiliser}. Moreover, most such approaches give no information about what states are easier to learn in comparison to others. Here, we propose a new generative energy-based representation of quantum many-body states derived from Gibbs distributions used for modeling the thermal states of classical spin systems. Based on the prior information on a family of quantum states, the energy function can be specified by a small number of parameters using an explicit low-degree polynomial or a generic parametric family such as neural nets, and can naturally include the known symmetries of the system. Our results show that such a representation can be efficiently learned from data using exact algorithms in a form that enables the prediction of expectation values of physical observables. Importantly, the structure of the learned energy function provides a natural explanation for the hardness of learning for a given class of quantum states.
\end{abstract}
\maketitle

\section*{Introduction}
Learning a generative model from several copies of a quantum state is quickly becoming an important task due to the ever-increasing size of quantum systems that can be prepared using various quantum information processing devices \cite{eisert2020quantum, torlai2020machine}. The quantum state with $n$ qubits is fully characterized by the density matrix $\rho \in \mathbb{C}^{2^n \times 2^n}$ with $2^{2n}-1$ parameters. Due to the exponentially increasing number of parameters with the system size, directly learning a quantum density matrix becomes rapidly impractical \cite{o2016efficient,haah2016sample}  except in special cases, such as Matrix Product  States (MPS) \cite{cramer2010efficient,schollwock2011density}, where hand-tailored quantum tomography techniques exploiting the known structure of the state can improve the representation efficiency \cite{cramer2010efficient, gross2010quantum, lanyon2017efficient, rocchetto2018stabiliser}. Alternatively, some compact representations for states can be hard to learn or can  be inefficient for the purposes of estimating observables.  For example,  the thermal state of a local quantum Hamiltonian can be specified using polynomially few parameters that encode the state's structure. However, this representation is not very useful in practice because currently no computationally efficient algorithms to learn the quantum Hamiltonian parameters are known, except in special cases \cite{haah2021optimal, anshu2020sample,rouze2021learning, onorati2023efficient}. Moreover, even if the quantum Hamiltonian is known, estimating the observables is in general difficult due to issues like the sign-problem \cite{sandvik1997introduction}. It is worth mentioning that if the aim of the tomographic technique is not to learn a generative model but to estimate a fixed subset of observables, then recent advances in \textit{shadow tomography} can be used in many practically interesting settings \cite{huang2020predicting,elben2022randomized}.

\begin{figure*}[!htb]
    \includegraphics{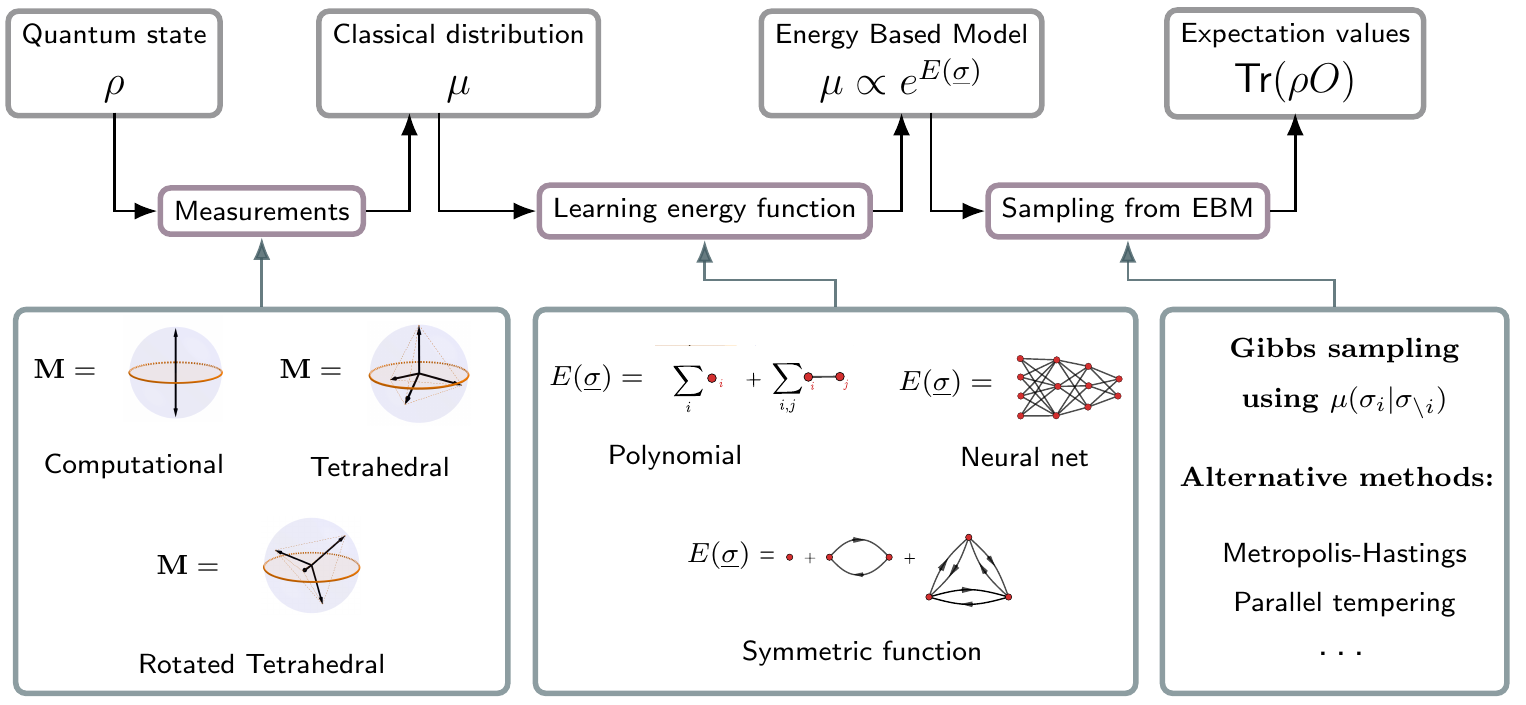}
    \caption{{\bf Summary of our approach to learning of energy-based representations for quantum states.} Our framework contains three modules in which different choices are possible. First, the choice of POVM defines a mapping from the quantum state to the corresponding classical representation. Second, the choice of the parametric family is related to the complexity of the energy function, as well as to the choice of the respective learning algorithm from the family of Interaction Screening estimators. Finally, once the energy function is learned, a suitable sampling algorithm can be used to generate samples and estimate observation values using the conditional probabilities obtained during the learning process. }
    \label{fig:cartoon2}
\end{figure*}

In many cases,  rather than trying to represent the quantum state directly, a more fruitful approach would be to learn a classical generative model for the measurement statistics generated by a quantum state. Since a state can be completely specified by the distribution of its measurement outcomes in an appropriately chosen basis, such a classical model can be used as an equivalent model for the quantum state. Most commonly, the  mapping of a quantum state to a classical distribution is performed using multi-qubit \emph{Positive Operator Valued Measures} (POVMs) in a form of a tensor product of informationally-complete single-qubit POVMs \cite{renes2004symmetric, carrasquilla2019reconstructing}. These POVMs are specified using operators $\mb{M}_{\sigma_i}$ for each qubit $i$, where $\sigma_i \in \{1, \ldots q\}$ is a classical variable that takes one of the $q$ states. For a quantum state on $n$ qubits, such a mapping generates a classical $n$-body distribution on classical spin configurations $\sigmab =  (\sigma_1, \ldots, \sigma_n)$: 
\beq \label{eq:POVM_prob_dist}
\mu(\sigmab) = \text{Tr}(\rho ~ \mb{M}^{(1)}_{\sigma_1} \otimes \mb{M}^{(2)}_{\sigma_2} \otimes \ldots \otimes \mb{M}^{(n)}_{\sigma_n}),
\eeq
which acts as an unambiguous representation of the quantum  state $\rho$. Non-informationally-complete POVMs can also be used to produce a reduced set of measurements, in which case $\mu(\sigmab)$ models only partial information on the quantum state. Given $\mu(\sigmab)$, the expectation value of any observable with respect to $\rho$ (or a reduced representation of $\rho$) can be computed as long as it is possible to sample from $\mu(\sigmab)$, see see \emph{Supplementary Information (SI)}, section \ref{app:POVM_sectionA}. These ideas have been exploited in recent works, where machine learning techniques originally developed to represent classical distributions have been used to learn representations for quantum many-body systems. A popular approach, originally introduced by Carleo and Troyer \cite{carleo2017solving}, is to model the amplitude and phases of a pure state using separate models. In an attempt to extend this representation to model practically abundant mixed states, recent works used different approaches such as introducing new degrees of freedom to purify the state \cite{torlai2018latent}, exploiting low-rank properties \cite{melkani2020eigenstate}, or using matrix factorization \cite{vicentini2022positive}. These methods often lead to black-box models for quantum states which makes it difficult to impose prior information about the state, such as locality properties or symmetries. A natural approach to modeling mixed states was used by Carrasquilla et al. \cite{carrasquilla2019reconstructing}, who directly used measurement statistics from a quantum state to construct the respective probabilistic representation using POVMs. This approach allows for the direct use of generative modeling techniques for learning quantum states. 

The major open question in representation learning of a classical distribution describing measurement data remains the lack of characterization of which states are hard or easy to model with standard generative modeling techniques. This makes choosing the right method for a given class of states an exercise in trial and error. Moreover, whereas many modern machine learning approaches rely on the generalization properties of neural networks to find the sparsity structure in the probability density, such a structure may only be apparent at the level of the energy function, as it is for instance the case for the thermal states of local Hamiltonians. In this work, we aim to rectify some of the common shortcomings encountered in a direct application of machine learning techniques to generative modeling of quantum states. We achieve this by modeling a distribution on statistics of measurement outcomes as an \emph{energy-based model} (EBM), 
\begin{equation}\label{eq:gibbs_dist}
\mu(\sigmab) = \frac{1}{Z} e^{E(\sigmab)},
\end{equation}
and by focusing on learning the general real-valued energy function $E(\sigmab)$ instead of the density itself. This approach is inspired by the power of Gibbs distributions used in modeling thermal states of classical many-body systems, where a simple quadratic energy function in the microscopic degrees of freedom induces distributions with highly non-trivial spin-glass structure \cite{mezard1984replica}. In applications beyond modeling classical thermal systems, EBMs have seen a recent resurgence in the generative modeling literature due to their simplicity and state-of-the-art generalization abilities on real-world datasets \cite{xie2016theory, kim2016deep, du2019energy, kumar2019maximum}.  In what follows, we leverage recent progress in rigorous learning of Gibbs distributions \cite{vuffray2019efficient} to identify the energy function for various families of pure and mixed quantum states, and to get insights on the complexity of representation, both from the perspective of learning and generating predictions, from the structure of the learned energy function and its effective temperature.

Our approach is schematically summarized in \figurename~\ref{fig:cartoon2}.  Our aim is to learn an EBM for the distribution $\mu(\sigmab)$, given $m$ measurement outcomes (i.e. samples from $\mu$) obtained by measuring independent copies of an $n$-qubit quantum state $\rho$  using a particular POVM.  The particular choice of the POVM determines the nature of the classical representation $\mu(\sigmab)$. In this work, we use computational, tetrahedral, and rotated tetrahedral POVMs, see \emph{SI}, section \ref{app:POVM_intro} for a precise definition. By studying the properties of $\mu(\sigmab)$, we can make an appropriate choice for the parametric family to represent the energy function $E(\sigmab)$. In an absence of any prior information, we use a generic neural-net parametric family to discover a sparse non-polynomial representation for the energy function \cite{abhijith2020learning}. When the structure is given by a low-order polynomial or some prior information on the state's symmetries is available, we use more specialized representations such as polynomial or symmetric function families, respectively. Given a specific parametric family, we use a state-of-the-art computationally- and sample-efficient method known as \emph{Interaction Screening} (IS) \cite{vuffray2016interaction, lokhov2018optimal, vuffray2019efficient, abhijith2020learning} to learn the parameters of the energy function. This learning procedure bypasses the intractability of the maximum likelihood and provides access to the conditional probabilities $\mu(\sigma_i | \sigmab_{\setminus i})$ for each spin $i$, which can then be used by a Markov Chain Monte Carlo (MCMC) method to effectively generate samples from the learned EBM and to estimate any expectation value the POVM gives us access to \cite{mackay2003information,abhijith2020learning}. In this work, we use Gibbs sampling for estimating observables from the learned model, although other methods can be utilized depending on the application focus \cite{katzgraber2006universality, schwing2011distributed, liu2015universal}. Our methodology naturally satisfies the two main desirable conditions for the generative modeling of quantum states: effective and practical algorithms for learning a classical representation; and the ability to infer quantities of interest from that learned representation. Finally, the structural information from the inferred effective energy function, such as the intensity of parameters and sparsity of interactions, can be directly connected to the information-theoretic complexity of the resulting representation \cite{santhanam2012information}, see \emph{SI}, section \ref{app:EBM_intro} for a detailed discussion on sample complexity scaling with the model parameters. Technical details regarding the IS method can be found in \emph{SI}, section \ref{app:GM}. Various parametric function families that are used for modeling the energy function are discussed in \emph{SI}, section \ref{app:FFamily}.

The main contributions of this work are as follows. We find that for many classes of quantum states, including thermal and ground states of local Hamiltonians, the choice of the correct function family to use to learn the EBM can be effectively made by studying systems on a small number of qubits. We can tractably learn these smaller systems in the infinite sample limit $(m \rightarrow \infty)$ as explained in the \emph{SI}, sections \ref{app:EBM_intro} and \ref{app:Methodology}, to understand the nature of the exact EBM representation for the state, and use this knowledge to inform our choices for larger systems, see \emph{SI}, section \ref{app:FFamily}.
For a given choice of the parametric function family modeling the energy function, we show how the effective temperature emerging from the learning procedure serves as a metric that quantifies the hardness of learning of different classes of quantum states. This is natural because the inverse temperature is the leading parameter that enters in the information-theoretic bounds for learning of EBMs, see \emph{SI}, section \ref{app:EBM_intro}.  
Finally, we showcase the advantages of ansatze that use a low number of parameters for representing quantum states. In all cases, we find that the EBM approach is well suited to learning classical generative models for quantum states and can be used to accurately estimate relevant observables. Our scaling experiments show that these energy-based methods are suitable for learning representations for large quantum systems, see \emph{SI}, section \ref{app:Additional_numerics}. As a particular example, we find that a low-parameter symmetric function ansatz for learning the energy function helps us perform better in fidelity estimation tasks when compared to other neural net-based methods \cite{huang2020predicting}, see \emph{SI}, section \ref{app:Fidelity} for precise definitions.

\section*{Results}

\subsection*{Learning mixed states.}
We start by exploring learning of energy-based representations for the mixed states. Specifically, we focus on the thermal states of the form
\begin{equation}
    \rho = \frac{\exp(-\beta H)}{\text{Tr}( \exp(-\beta H))},
\end{equation}
where $H$ is a local Hamiltonian. For the illustrations in this section, we use the transverse field Ising model (TIM) family of Hamiltonians with a uniform transverse field, $H_{TIM} =  \sum_{i<j} J_{ij} Z_i Z_j + g \sum_i X_i$.  
These models are particularly useful for studying properties of the learning algorithms as they allow for easy generation of samples, either using Matrix Product States (MPS) in 1D \cite{schollwock2011density, itensor} or using Quantum Monte-Carlo (QMC) methods in higher dimensions owing to the absence of the sign-problem in these models \cite{sandvik1997introduction}.

\begin{figure*}[!htb]
\begin{subfigure}[t]{\textwidth}
    \includegraphics{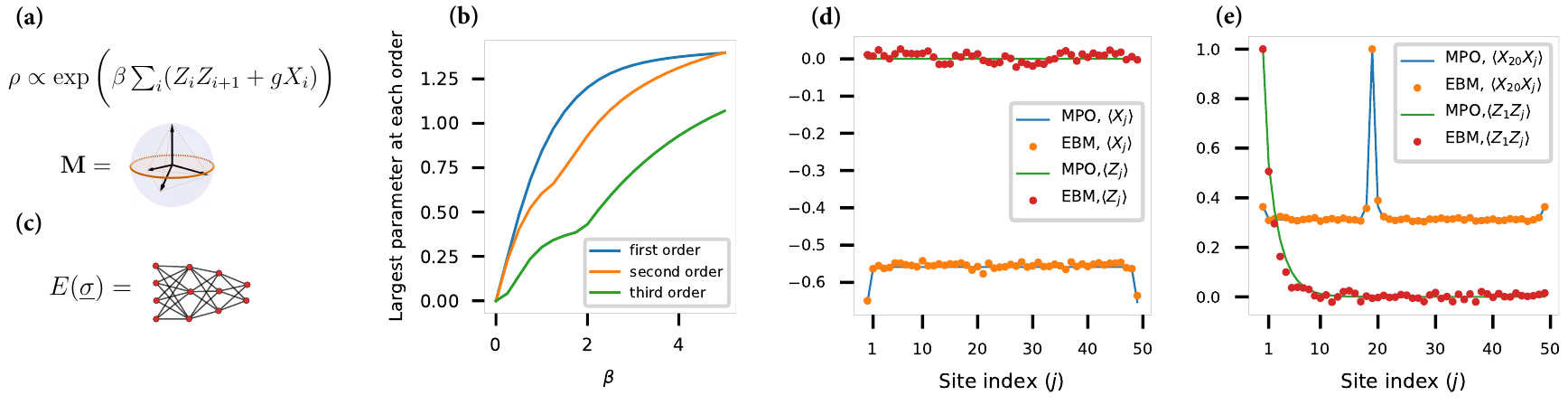}
\end{subfigure}
\begin{subfigure}[t]{\textwidth}
    \includegraphics{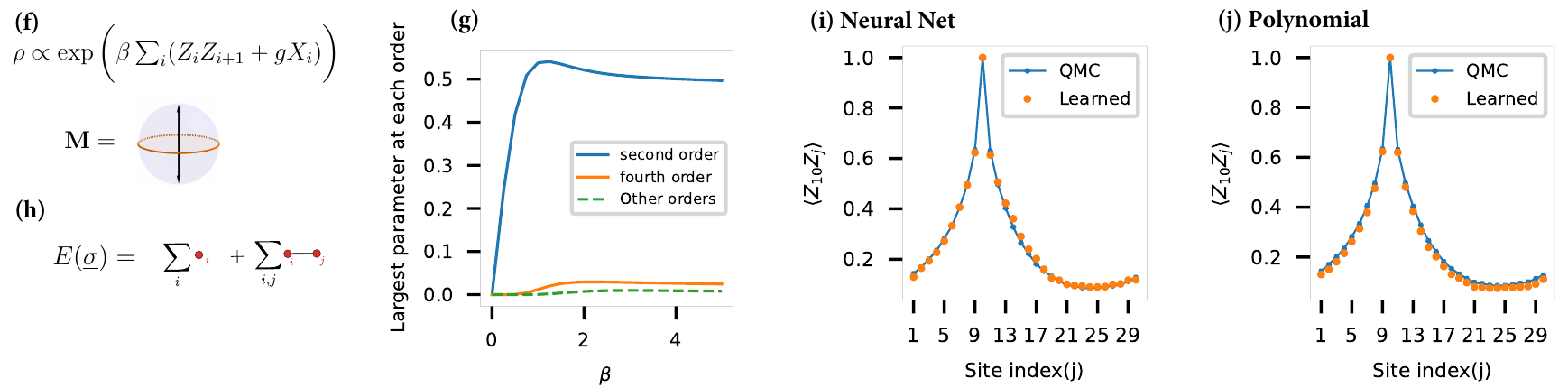}
\end{subfigure}
 \resizebox{!}{0.0cm}{\begin{minipage}{\textwidth}
 \begin{tabular}[t]{c c c c}
 \begin{tabular}[t]{c}
        \smallskip
            \begin{subfigure}[t]{0.22\textwidth}
            \caption{}
            \vspace{0.1in}
                \centering
                \includegraphics[scale=0.65]{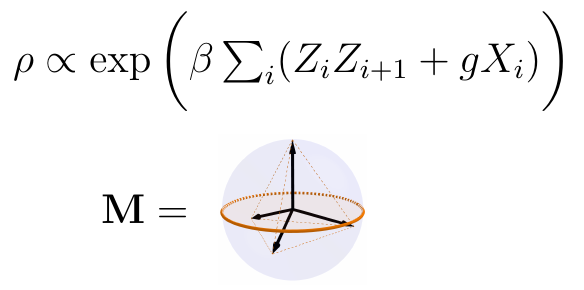}
                \label{fig:2a}
            \end{subfigure}\\
            \begin{subfigure}[t]{0.22\textwidth}
            \addtocounter{subfigure}{1}
                \caption{}
                \centering
                \vspace{-0.3in}
               \hspace{-0.2in} \includegraphics[scale=0.7]{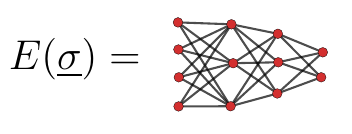}
            \end{subfigure}
\end{tabular}
\begin{subfigure}[t]{0.25\textwidth}
            \addtocounter{subfigure}{-2}
    \caption{}
    \vspace{-0.06in}
    \includegraphics[width=\textwidth]{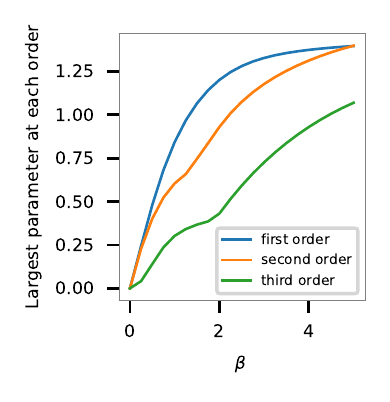}
    \label{fig:order_gibbs_POVM}
\end{subfigure}  
\begin{subfigure}[t]{0.25\textwidth}
    \setcounter{subfigure}{3}
    \caption{}
    \includegraphics[width=\textwidth]{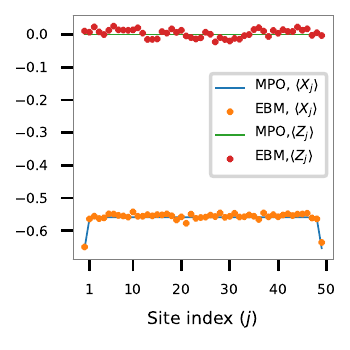}
    \label{fig:2c}
\end{subfigure}    
\begin{subfigure}[t]{0.25\textwidth}
    \caption{}
    \includegraphics[width=\textwidth]{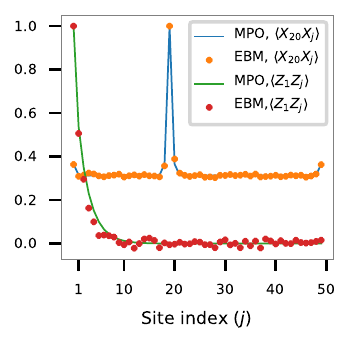}
    \label{fig:2d}
\end{subfigure}\\
\end{tabular}
 \begin{tabular}[t]{c c c c} \begin{tabular}[t]{c}
        \smallskip
            \begin{subfigure}[t]{0.22\textwidth}
            \caption{}
            \vspace{0.1in}
                \centering
                \includegraphics[scale=0.65]{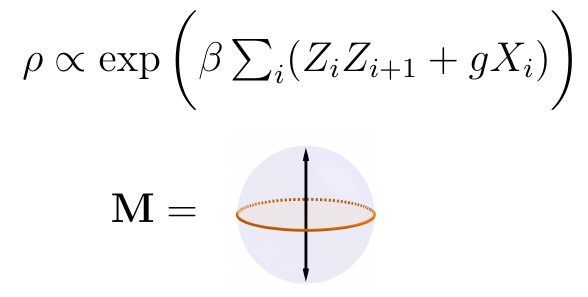}
                \label{fig:2e}
            \end{subfigure}\\
            \begin{subfigure}[t]{0.22\textwidth}
            \addtocounter{subfigure}{1}
                \caption{}
                \centering
                \vspace{-0.3in}
               \hspace{0.1in} \includegraphics[scale=0.7]{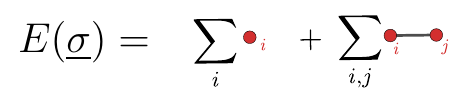}
            \end{subfigure}
\end{tabular}
\begin{subfigure}[t]{0.25\textwidth}
            \addtocounter{subfigure}{-2}
\caption{}
\includegraphics[width=\textwidth]{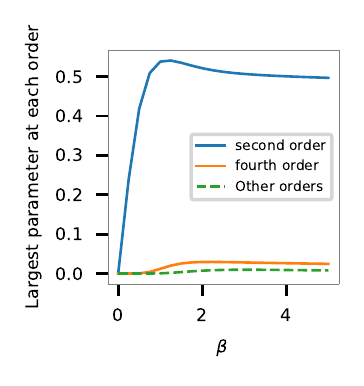}
\label{fig:order_gibbs}
\end{subfigure}
\begin{subfigure}[t]{0.25\textwidth}
            \addtocounter{subfigure}{1}
    \caption{Neural Net}
    \vspace{0.13in}
    \includegraphics[width=\textwidth]{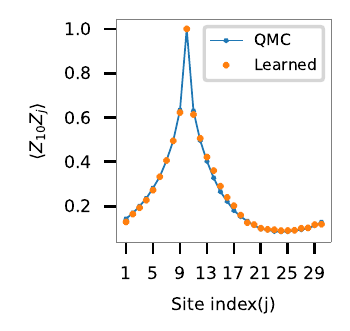}
\label{fig:Corr_1D_TIM_NN}
\end{subfigure}    
\begin{subfigure}[t]{0.25\textwidth}
    \caption{Polynomial}
    \vspace{0.13in}
    \includegraphics[width=\textwidth]{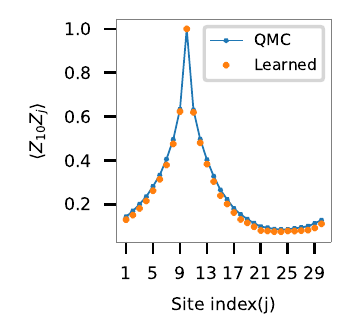}
\label{fig:Corr_1D_TIM_poly}
\end{subfigure}
\end{tabular}
 \end{minipage}}
\caption{ {\bf Learning classical energy-based representations for quantum thermal states.} {\bf (a)} EBM learned from the TIM thermal state measured using the tetrahedral POVM. {\bf (b)} The strength of interactions at each order for a third-order polynomial energy function exactly representing the thermal state of a $5$-qubit TIM on a line measured using the Tetrahedral POVM. Here the absolute value of the largest coefficient in the polynomial representation at each order is plotted against the inverse temperature. For these experiments, we fix $g=1$. We see that terms at all orders in the energy function are not negligible. {\bf (c)}  A neural net representation is most appropriate given the presence of higher-order terms in the polynomial representation. {\bf (d)} One- and {\bf (e)} two-body expectation values inferred from the neural net representation for the ferromagnetic 1D TIM at $\beta = 1, g = 1,$ with open boundary conditions. The EBM representation is learned using $m = 10^5$ samples, and observables are computed by drawing $4\times 10^5$ samples from the learned model.
{\bf (f)} TIM thermal state measured in the computational basis only. {\bf (g)} Strength of interactions at each order for an EBM exactly representing the thermal state of a 5-qubit ferromagnetic TIM using measurements in the computational basis.
We see that the second-order terms dominate over other terms. {\bf (h)} This means that the energy function can be well approximated using a second-order polynomial. $ZZ$  expectations estimated from EBMs represented using {\bf (i)} neural nets and {\bf (j)} quadratic polynomials, respectively. Both representations are learned for a $30$ spin thermal TIM at $\beta = 2, g = 1$ with $m = 10^6$ samples, with periodic boundary conditions. The neural net model has $28380$ trainable parameters while the polynomial model has only $900.$ Further details about the numerical experiments are given in \emph{SI}, section \ref{app:Methodology}. }
\label{fig:gibbs_tetra_corr}
\end{figure*}

First, we consider the TIM Hamiltonian on a 1D lattice with $J_{i,i+1} = -1 \, \forall i$ and zero otherwise, measured in the tetrahedral POVM, see \figurename~\ref{fig:2a}. We can obtain an insight on whether a more specific polynomial function family (rather than the generic neural-net one) can effectively capture the distribution generated by a quantum state and POVM combination using the following approach: for a small state, \emph{exact} learning of the energy function in the polynomial basis would give us valuable information about the types of terms present in the energy function and their symmetry properties, as explained in the \emph{SI}, sections \ref{app:EBM_intro} and \ref{app:Methodology}.
In \figurename~\ref{fig:order_gibbs_POVM}, we generate the exact distribution defined by \eqref{eq:POVM_prob_dist} for a $5$ qubit thermal state using matrix exponentiation and learn an EBM for it using the polynomial family with up to third-order terms. We see that the polynomial representation learned here has significant terms at all orders including order three, which is in contrast with the respective quantum Hamiltonian which only had second-order interactions. This experiment suggests that using the polynomial representation is not very efficient for this state/POVM combination as the computational complexity of the learning procedure will scale as $\Theta(n^L)$ for a $n$-qubit state and for higher-order terms of order $L$. This motivates the use of a fully-connected neural net as the parametric function family for the energy function representation.

The specific algorithm in the Interaction Screening suite of methods adapted to the neural-net parametric family is known as \emph{NeuRISE} (see \emph{SI}, section \ref{app:GM} for more details), which has been shown to be able to learn energy functions with higher order terms in an implicit sense with fewer parameters when the complexity of learning using polynomials scales unfavorably due to the presence of higher order terms \cite{abhijith2020learning}. In Figs.~\ref{fig:2c} and \ref{fig:2d}, we study the ferromagnetic 1D TIM with $50$ qubits  in the quantum critical phase \cite{sachdev_2011}. We compare the one- and two-body correlations inferred from the learned classical representation to the ones generated from a Matrix Product Operator (MPO) representation of the state found by imaginary time evolution using the ITensor library \cite{itensor}, in Figs.~\ref{fig:2c} and \ref{fig:2d}, respectively. We see that the EBM representation is able to reproduce expectation values of both diagonal and off-diagonal observables.

The nature of the EBM representation can change considerably depending on the POVM used. We demonstrate this using the same thermal ferromagnetic TIM state, now measured in the computational basis rather than using the tetrahedral POVM, see \figurename~\ref{fig:2e}. Note that for the general states, measurements in the computational basis do not provide an informationally complete set sufficient for unambiguously specifying the entire state. \figurename~\ref{fig:order_gibbs} shows results for the exact learning on a thermal state of a thermal ferromagnetic TIM state from measurements in the computational basis. The results are markedly different from \figurename~\ref{fig:order_gibbs_POVM}: we see that the second-order interactions dominate the energy representation, and hence the energy function can be well approximated by a quadratic polynomial. In \figurename~\ref{fig:Corr_1D_TIM_NN} and \ref{fig:Corr_1D_TIM_poly}, we show that, similarly to the case of the tetrahedral POVM, the correlation functions are well predicted by both general neural-net and specialized polynomial parametric representations of the energy function in the computational basis. But the polynomial representation uses significantly fewer parameters and is ideal for use in a resource-limited setting.

Results showing scaling of error with the size of the system and number of samples for the TIM are given in \figurename ~\ref{fig:scaling_gibbs_2D} in \emph{SI}, Section \ref{app:Additional_numerics}. Specifically, \figurename~\ref{fig:samples_vs_err_gibbs_2D} and \figurename~\ref{fig:samples_vs_err_gibbs_2D_critical} show the average error in $ZZ$ expectation values for the two different values of $g$, with $\beta=1$  for a 2D TIM. Here it is observed that the learning algorithm performs better at higher values of $g$. This behavior is expected, as for $g \rightarrow \infty$ the learned distribution gets closer to a high-temperature Gibbs distribution in the computational basis, and it is known from prior works that learning high-temperature EBMs are easy \cite{vuffray2016interaction, lokhov2018optimal, vuffray2019efficient}. On the other hand, for $g \rightarrow 0$  we are close to the $\beta=1$ thermal state of a classical Ising model. Learning lower-temperature classical Gibbs distributions is known to be difficult in general from sample-complexity lower bounds \cite{santhanam2012information}. The results for a fixed $\beta$ suggest that the value of $g$ plays an important role in determining the final effective temperature of the distribution we are learning, with larger values of $g$ producing classical distributions with a higher effective temperature. In the \emph{SI}, section \ref{app:Additional_numerics} we also observe that the error does not increase considerably with the number of qubits in the system. This is consistent with  the logarithmic scaling of sample complexity with the number of spins observed for learning EBMs. \cite{vuffray2016interaction, lokhov2018optimal, vuffray2019efficient}.

\subsection*{Learning ground states.}

\begin{figure*}
    \centering
    \includegraphics{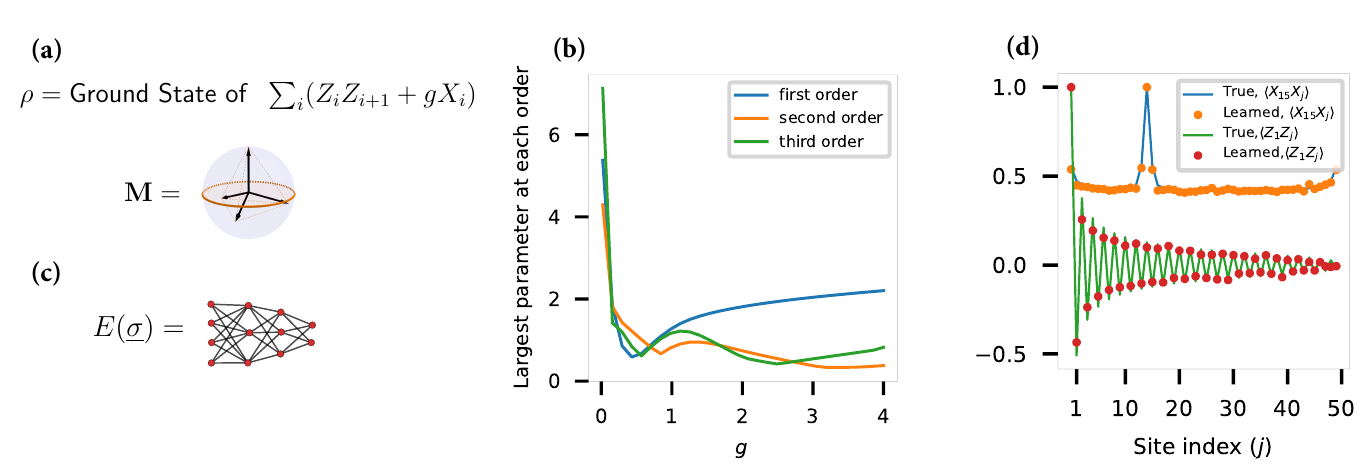}
     \resizebox{!}{0.0cm}{\begin{minipage}{\textwidth}
\begin{tabular}[t]{c c c}
 \begin{tabular}[t]{c}
        \smallskip
            \begin{subfigure}[t]{0.22\textwidth}
            \caption{}
            \vspace{0.1in}
                \centering
                \includegraphics[scale=0.65]{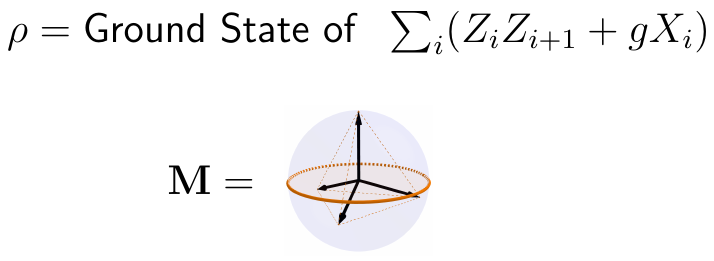}
                \label{fig:3a}
            \end{subfigure}\\
            \begin{subfigure}[t]{0.22\textwidth}
            \addtocounter{subfigure}{1}
                \caption{}
                \centering
                \vspace{-0.3in}
                \includegraphics[scale=0.7]{Fig/Efn_nn.pdf}
            \end{subfigure}
\end{tabular}
\qquad
    \begin{subfigure}[t]{0.25\textwidth}
            \addtocounter{subfigure}{-2}
    \caption{}
    \vspace{-0.05in}
    \includegraphics[width=\textwidth]{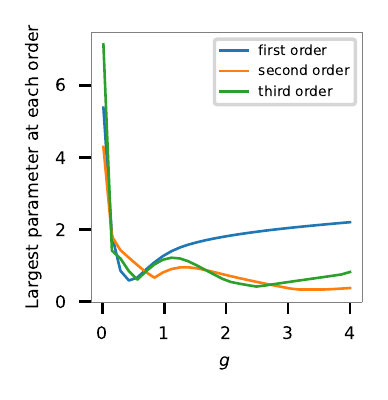}
    \label{fig:order_ground}
    \end{subfigure}
\begin{subfigure}[t]{0.25\textwidth}
            \addtocounter{subfigure}{1}
      \caption{}
    \includegraphics[width=\textwidth]{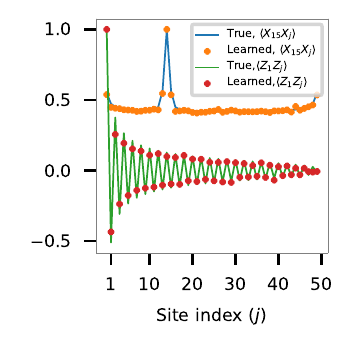}
\label{fig:Corr_1D_TIM_ground}
\end{subfigure}
\end{tabular}\\
\end{minipage}}
    \caption{ {\bf Learning classical energy-based representations for ground states of quantum systems.} {\bf (a)} Here, we study EBM representation of a ground state of the anti-ferromagnetic TIM measured in the Tetrahedral basis. {\bf (b)} Strength of interactions at each order for an EBM exactly learned from the ground state of a $6$-qubit anti-ferromagnetic TIM on a line. The figure shows that a second-order representation cannot capture all the features of this model. The relative importance of all orders is especially pronounced when the Ising term dominates the transverse field. {\bf (c)} Because of these higher order terms, we find that a neural net ansatz is necessary to capture all the features of the model  {\bf (d)} Two-body expectation values for a $50$-spin model at $g=1$. The EBM learned from this state is able to accurately learn the long-range behavior of the ground state at the critical point. All results are from a neural net ansatz  are trained from samples drawn from an MPS representation of the ground state, found using DMRG \cite{itensor, schollwock2011density}. The EBM was learned using $m = 10^5$ training samples. Further details about the numerical experiments are given in \emph{SI}, section \ref{app:Methodology}. }
\end{figure*}

Learning a distribution of classical ground states is  hard due to the high sample complexity of methods at lower temperatures. However, the classical distribution of ground states of quantum systems modeled as EBM, have an effective temperature depending on the strength of off-diagonal terms in the Hamiltonian. Our results below show that EBMs learned from the zero temperature states of quantum Hamiltonians exhibit a certain effective temperature property that can aid the learning process. As an illustration, we consider the problem of learning an informationally-complete representation of the ground state of an anti-ferromagnetic TIM from samples obtained from measurements in the tetrahedral POVM, see \figurename~\ref{fig:3a}.

In actuality, as the TIM Hamiltonian is stoquastic, the ground state can be fully specified by computational basis measurements  \cite{bravyi2008complexity, bravyi2015monte}. \figurename~\ref{fig:order_gibbs} shows that in the regime $\beta \to \infty$ a polynomial representation for the ground state of the TIM learned from measurements in the $Z$ direction alone will have predominantly second-order terms.
But in general, without prior information about the Hamiltonian, it is hard to determine whether a ground state can be fully specified by only computational basis measurements. So in these experiments, we deliberately do not use the fact that the TIM Hamiltonian is stoquastic and attempt to learn  a more general representation of the ground state using the tetrahedral POVM.

In \figurename~\ref{fig:order_ground}, we use our strategy of learning an exact distribution for a small system of size $n=6$ to get an insight into the structure of the energy function. \figurename~\ref{fig:order_ground} shows the strength of terms present in an energy function learned using a third-order polynomial ansatz. Notice that the effective temperature of the EBM is inversely proportional to the strength of interactions shown in \figurename~\ref{fig:order_ground}.  We see that in the small $g$ regime, the strength of interactions blows up as $g \rightarrow 0$, thus providing evidence for the low effective temperature of $\mu(\sigmab)$ at low values of $g$. In the large $g$ regime, on the contrary, the first-order terms are dominant and the effective temperature is not very large. This means that a polynomial model will be able to give a satisfactory approximation in the high $g$ regime. We see that below the critical point $(g=1)$, terms at all orders are equally important. This means that we will have to resort to a universal neural net ansatz due to the presence of higher-order terms in the energy function. 

The effect of $g$ on learning a $60$ qubit state from the same family is given in \figurename~\ref{fig:err_vs_g} in \emph{SI}, section~\ref{app:Additional_numerics}. In line with the results for thermal states in \figurename~\ref{fig:scaling_gibbs_2D}, we see that an EBM can easily be learned for states that possess paramagnetic order. But as expected, the sample complexity of the learning procedure increases when the states come close to the ground state of the classical model. Therefore, the observed effect of $g$ on learning is much more pronounced for ground states when compared to thermal states. 

The results of using NeurISE for an anti-ferromagnetic TIM on a 50 qubit 1D lattice with open boundaries at the critical point are given in \figurename~\ref{fig:Corr_1D_TIM_ground}. We see that this method has no problems learning states that possess long-range order. This observation is in line with what has been established in classical models as well \cite{lokhov2018optimal}.

Similarly to learning representations of thermal states, our experiments with ground states also show that the error in the learning procedure scales favorably with the system size. In \emph{SI}, section \ref{app:Additional_numerics}, we give scaling results for learning ground states, and discuss learning EBM representations of ground states for the Heisenberg model in 2D.

\subsection*{Learning states with symmetries.}
Another important aspect of quantum states that can greatly aid the learning process is the presence of symmetries. Symmetries can significantly reduce the size of the hypothesis space that a machine-learning method must optimize over. But it is not always easy to build an ansatz that respects the known symmetries of a model \cite{barnard1991invariance, cohen2018spherical,schutt2017schnet}.
The most common symmetry encountered in physical systems is translational invariance, which can be incorporated into both the energy function representation and the IS learning method rather easily. Once the state is represented by a classical distribution $\mu(\sigmab)$ using the relation \eqref{eq:POVM_prob_dist}, this distribution inherits the translational symmetry of the state. This in turn implies that the conditional distribution $\mu(\sigma_i | \sigma_{\setminus i})$ of every variable $i$ in the EBM is the same. And since the EBM is learned via these conditionals, the learning step needs only be done for one variable. This leads to a factor $n$ reduction in the computational cost of learning and nothing extra has to be built into the energy function ansatz that we use.
For other types of symmetries, the parametric function family we use has to be appropriately chosen to respect the model's symmetries
incorporated into the EBM. This makes learning the representation significantly more efficient due to a reduced number of parameters. 

As an example of this approach, we will demonstrate learning states that are invariant under any permutation of qubits. The simplest way to use this prior information is to use a fully symmetric parametric function family to represent the energy function. Any such function acting on $n$ discrete variables where each variable takes $q$ different values can be specified using only $O(n^{q-1})$ parameters. This is because the value of a fully symmetric function can only depend on the number of occurrences of each of the $q$ possible values. For an $n$-qubit state with the tetrahedral POVM, this implies a representation of size $O(n^3)$, for the most general fully symmetric energy function. We can also use a neural net that respects this permutation symmetry here. But with a neural net, there is no guarantee that every symmetric function can be represented using a comparable number of parameters. By specifying the most general symmetric function we can guarantee that the exact EBM that represents the state in question is always considered by the algorithm. Moreover, the optimization problem here is convex in the model parameters compared to the non-convex optimization required to learn a neural net \cite{abhijith2020learning}. 

To illustrate learning of permutation-symmetric quantum states, we consider learning a classical EBM representation for the GHZ states:
\begin{equation}
    \ket{GHZ^n_{\pm}} = \frac{ \ket{0}^{\otimes n}  \pm \ket{1}^{\otimes n}}{\sqrt{2}}
\end{equation}

\begin{figure*}
    \includegraphics{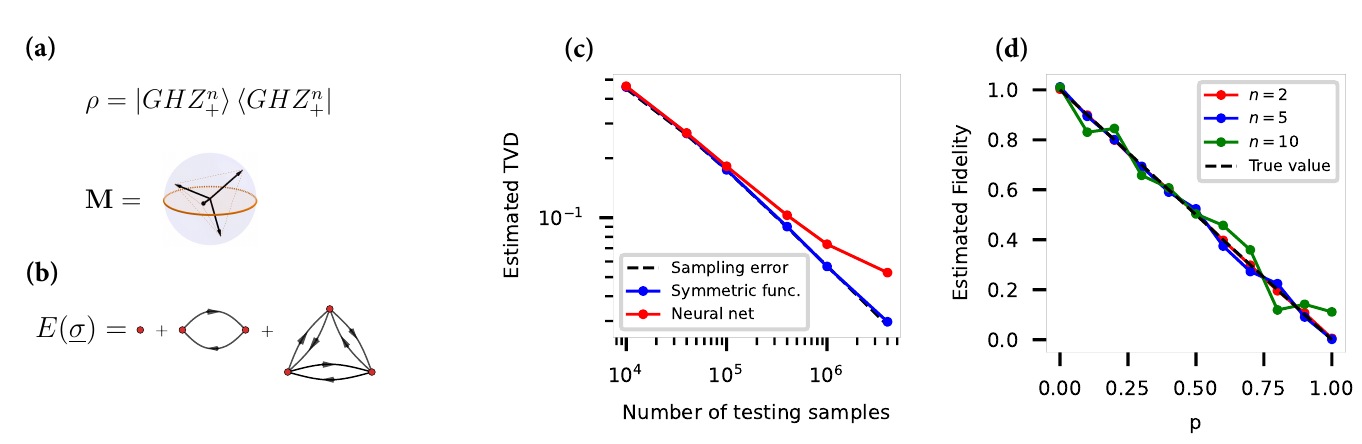}
     \resizebox{!}{0.0cm}{\begin{minipage}{\textwidth}
      \begin{tabular}[t]{c c c}
 \begin{tabular}[t]{c}
            \begin{subfigure}[t]{0.22\textwidth}
            \caption{}
            \vspace{0.1in}
                \centering
                \includegraphics[scale=0.65]{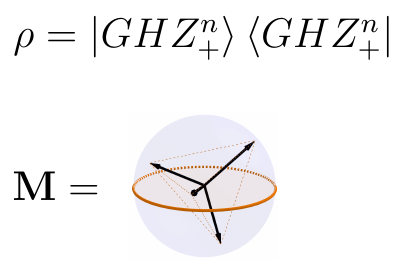}
            \end{subfigure}\\
            \begin{subfigure}[t]{0.22\textwidth}
                \caption{}
                \centering
                \vspace{-0.2in}
                \includegraphics[scale=0.7]{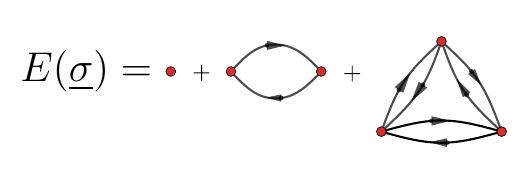}
            \end{subfigure}
\end{tabular}
\qquad
    \begin{subfigure}[t]{0.25\textwidth}
    \caption{}
    \includegraphics[width=\textwidth]{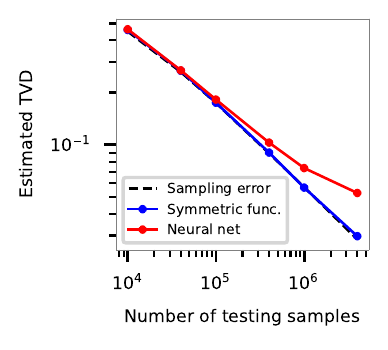}
    \label{fig:TV_GHZ_compare}
    \end{subfigure}
    \begin{subfigure}[t]{0.25\textwidth}
    \caption{}
    \includegraphics[width=\textwidth]{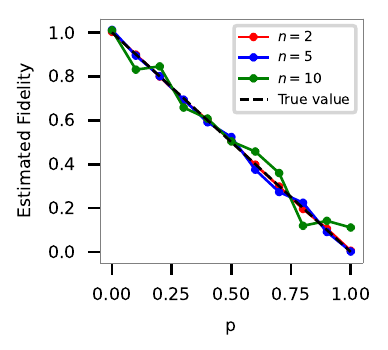}
    \label{fig:GHZ_prep}
    \end{subfigure}
    \end{tabular}
    \end{minipage}}
    \caption{ {\bf Learning classical energy-based representations for permutation invariant systems. }{\bf (a)} We use the tetrahedral POVM rotated by a fixed one-qubits Haar random unitary for learning GHZ states. {\bf (b)} Given prior information about the symmetry of the state, we can use a fully permutation invariant function in the IS method. {\bf (c)} The Total Variation Distance (TVD) of the learned distribution from the true distribution obtained by measuring a $7$ qubit GHZ state in the tetrahedral POVM. The $x$-axis here is the number of samples drawn from a learned model to estimate the TVD. The dotted line is the sampling error, which is TVD estimated from the exact classical representation of the state given by the POVM. We see that the symmetric function ansatz performs much better than neural-net ansatz learned with NeurISE \cite{abhijith2020learning} without any symmetry assumptions. Both models are learned using $10^6$ state measurements. The neural net has $1043$ trainable parameters while the symmetric function has $840$  parameters {\bf (d)} Comparison of the fidelity between $\rho_{n,p}$
    for $p\in[0,1]$ with $\ket{GHZ^n_+}$ estimated from a symmetric EBM  for $\rho_{n,p}$. The symmetric EBM  is able to produce the correct fidelity behavior here even when $p \rightarrow 1.$ It has been reported that for this benchmark, NNQS methods which compute a classical upper bound for quantum fidelity, are not able to reproduce the correct behavior when $p$ gets close to $1$ \cite{carrasquilla2019reconstructing, huang2020predicting}. All the models here are learned using $6\times 10^4$ samples. Details of fidelity estimation are provided in \emph{SI}, section \ref{app:Fidelity}.}
\end{figure*}

Using the exact learning procedure, we find that the EBM representation of these states in the tetrahedral basis has terms of all orders. This explains the results in Ref. \cite{carrasquilla2019reconstructing} where an RBM with a few hidden units was found incapable of representing this state when $n$ became large. In order to learn an energy-based representation we either have to resort to a generic neural-net parametrization, or incorporate knowledge about the state symmetries. The comparison of these two approaches is given in \figurename~\ref{fig:TV_GHZ_compare}, where we estimate the Total Variation Distance (TVD) between the distribution representing $\ket{GHZ^7_+}$ and learned models by drawing new samples from these models. We see that the symmetric ansatz outperforms a neural net ansatz of comparable size. After drawing $4\times10^6$ samples from these models learned from $m =10^6$ measurements, we find that the estimated TVD between the symmetric function ansatz and the true state is only $4\%$ more than the inherent finite-sampling error in the estimation procedure. While the TVD estimated for the neural net is $77\%$ more compared to the sampling error.

In an experiment involving symmetric states, first studied in \cite{huang2020predicting}, the fidelity of $\ket{GHZ^n_{+}}$ is estimated with respect to another mixed state of the form, $ \rho_{n,p} = (1-p) \ket{GHZ^n_+}\bra{ GHZ^n_+} + p \ket{GHZ^n_-}\bra{GHZ^n_-}$ for $p\in[0,1]$. It was been reported that for this mixed state example, generative models learned using the Neural Net Quantum States (NNQS) ansatz \cite{huang2020predicting, carrasquilla2019reconstructing} struggled to find the correct fidelity as $p$ got closer to $1$. The results of applying our method on this problem with a symmetric function ansatz are given in \figurename~\ref{fig:GHZ_prep}. Compared to the NNQS results reported in \cite{huang2020predicting}, we see that the IS method with the symmetric function ansatz is able to better predict the overlap of the above state with the $\ket{GHZ^n_+}$. 

\section*{Discussion}

We have presented a novel approach for learning generative models for quantum states. The classical energy based representation used here is flexible enough to incorporate prior information about the state including its symmetries. Using a few representative cases of pure or mixed quantum states, we have illustrated the respective advantages of using different parametric families for modeling energy functions of the emerging classical representations. We have also observed that the classical energy function inherits a certain effective temperature from the quantum states. This effective temperature acts as an information-theoretic metric for the hardness of learning of a classical representation for a given class of quantum states.

There are many interesting directions to explore using the classical energy-based representations of quantum states. An interesting theoretical question is whether the classical energy function inherits locality or low-degree structure from the quantum Hamiltonian. This possibility cannot be ruled out completely from the exact learning experiments presented in this work. If such a connection can be shown for the EBM for some class of  quantum states, then that implies an efficiently learnable representation for those states. Polynomial representations of energy functions are well suited for such a theoretical analysis when compared to black-box techniques that use neural nets.
Another important direction is learning from noise-corrupted data, as most real-world tomographic data is corrupted by errors in the measurement process \cite{maciejewski2020mitigation, geller2021toward}. One of the advantages of the IS-based learning methods is that they can be made robust to many types of noisy channels \cite{goel2019learning}. We leave the development of energy-based learning methods with robustness to measurement noise as future work.
\vspace{-0.2in}
\subsection*{Data availability}
Data used from the plots in this work is available from the corresponding author upon reasonable request.
\subsection*{Code availability}
Code to reproduce the experiments described in this paper can be found at \url{https://github.com/abhijithjlanl/EBM-Tomography}
\subsection*{Acknowledgements} The authors acknowledge support from the Laboratory Directed Research and Development program of Los Alamos National Laboratory under Projects 20210674ECR and 20230338ER. This research used computational resources provided by the Darwin testbed and the Institutional Computing program at LANL.
\subsection*{Author contributions}
All authors designed the research,
wrote the manuscript, reviewed and edited the paper. AJ wrote the code used in this paper.
\subsection*{Competing Interests}
The authors declare no competing interests.
\subsection*{Additional Information}
\textbf{Supplementary Information} is available for this paper.
\textbf{Correspondence and requests for materials} should
be addressed to AJ.

\bibliographystyle{naturemag}
\bibliography{main}
\balancecolsandclearpage

\newpage

\appendix

\onecolumngrid

\begin{center}
{\LARGE Supplementary Information}\\
\end{center}

\twocolumngrid
\paragraph*{\bf Notations:} Vector-valued functions are denoted by bold space (eg. $\bm{f}$, $\bm{\Phi}$). The boldface is removed while referring to  their individual outputs (eg.  $f_a$, $\Phi_a$). Vector variables are denoted by an underline ($\ul{\sigma}$). The underline is not used while referring to their individual outputs ($\sigma_u$). The subset of variables from this vector given by an index set $K$ is dented by $\sigmab_K$. The shorthand,  $\sigmab_{\setminus u}$ is used to denote all the elements of $\sigmab$ excluding $\sigma_u$. $\{X,Y,Z\}$ are reserved for single qubit Pauli operators.

\section{Probabilistic representation of quantum states by POVMs}
\label{app:POVM_sectionA}

For constructing classical representations of quantum states, we use a probability distribution that is generated by the quantum state when we perform a certain set of measurements on it. We use a \textit{Positive Valued Operator Measure} (POVM) to represent such a set of measurements \cite{renes2004symmetric, nielsen2002quantum, carrasquilla2019reconstructing}. POVMs are a set of positive operators such that they sum to the identity. A POVM with $q$ elements is defined as,
\begin{equation}\label{eq:POVM_defn}
   \mb{M}_\sigma \geq 0 ~~ \forall  ~\sigma \in [q] , ~~\sum_{\sigma}  \mb{M}_\sigma  = I.
\end{equation}

The POVM maps a quantum state $\rho$ to a probability vector according to the Born rule,
\begin{equation}
    \mu(\sigma) =  \text{Tr} ( \mb{M}_\sigma \rho ).
\end{equation}
 The two conditions in Eq. \eqref{eq:POVM_defn} is sufficient to ensure that the $\mu(\sigma)$ values form a valid probability distribution.
 
For many-body states whose Hilbert spaces have a tensor product structure POVMs can be constructed by taking tensor products of single-body POVMs. For an $n$ qubit system with a POVM of size $q$ for each qubit, the POVM elements for the full system will have the form,
 \begin{equation}\label{eq:many_body_povm}
\mb{M}_{\ul{\sigma}} =  \mb{M}^{(1)}_{\sigma_1} \otimes \mb{M}^{(2)}_{\sigma_2} \otimes \ldots \otimes \mb{M}^{(n)}_{\sigma_n}.
\end{equation}
 
Here $\ul{\sigma}$ is now a multi-index that runs over all the $q^n$ possibilities. This POVM naturally maps a quantum state over $n$ qubits to a high-dimensional probability distribution,
 \begin{equation}\label{eq:POVM_prob_dist_1}
      \mu(\ul{\sigma}) =  \text{Tr} ( \mb{M}_{\ul{\sigma}} \rho ).
 \end{equation}

Measuring the quantum state using the above POVM gives us samples drawn from this distribution. We can then use these samples to learn an EBM for the quantum state in question. 

\paragraph*{Informationally complete representation.}

A POVM is known as \textit{informationally complete} if the relation in Eq. \eqref{eq:POVM_prob_dist} can be inverted to unambiguously reconstruct the quantum state $\rho.$  This inversion can be performed using a set of dual operators defined by,
\begin{equation}
    \mb{D}_{\ul{\sigma}} = \sum_{\ul{\sigma}^\prime} [C^{-1}]_{\ul{\sigma}, \ul{\sigma}^\prime} \mb{M}_{\ul{\sigma}^\prime}, ~~C_{\ul{\sigma}, \ul{\sigma}^\prime} =  \text{Tr}(\mb{M}_{\ul{\sigma}}\mb{M}_{\ul{\sigma}^\prime}).
\end{equation}
If POVM operators are linearly independent, then the matrix $C$ is always invertible. On top of linear independence, if the POVM operators can span the entire operator space then the POVM is informationally complete. In this case, the density matrix can be reconstructed from the distribution of measurement outcomes using the dual operators,

\begin{equation}\label{eq:dual}
    \rho = \sum_{\ul{\sigma}} \mu(\ul{\sigma}) \mb{D}_{\ul{\sigma}}.
\end{equation}

If the POVM operators for an $n$-qubit system are formed by the tensor product of single qubit POVMs then both the $C$ matrix and the dual operators will inherit this tensor product structure. This property allows for the estimation of local observables and other quantities of interest from a model that can generate samples from $\mu$. If ${\ul{\tau}^{(1)}, \ldots \ul{\tau}^{(N)}}$ are such samples then an unbiased estimate for $\text{Tr}(\rho O)$ can be constructed for any $O.$
\begin{align}\label{eq:approx}
    \text{Tr}(\rho O) &= \underset{\ul{\tau} \sim \mu }{\mathbb{E}} \text{Tr}(\mathbf{D}_{\ul{\tau}} O  ) \\
    &\approx  \frac{1}{N} \sum_{k=1}^N \text{Tr}(\mathbf{D}_{\ul{\tau}^{(k)}} O  )  \\
    &= \frac{1}{N} \sum_{k=1}^N \text{Tr}( O ~ \mb{D}^{(1)}_{\tau^{(k)}_1} \otimes \mb{D}^{(2)}_{\tau^{(k)}_2} \otimes \ldots \otimes \mb{D}^{(n)}_{\tau^{(k)}_n}).
\end{align}

If the operator $O$ has an efficient representation as an MPO or as a sum of a few Pauli strings then this estimate can be computed efficiently \cite{carrasquilla2019reconstructing}. 

\section{Choice of POVMs}
\label{app:POVM_intro}

To map quantum states to probability distributions we mainly use the Tetrahedral POVM. This POVM corresponds to a $4$-outcome measurement and is informationally complete. The single qubit POVM operators are given by,
\begin{alignat}{2}
   \mathbf{M}_1 &= ~\begin{pmatrix}
\frac12 & 0 \\[0.2cm]
0 & 0 
\end{pmatrix},\quad \quad  
\mathbf{M}_2 = ~ \frac12\begin{pmatrix}
\frac13 & \frac{\sqrt2}{3} \\[0.2cm]
 \frac{\sqrt2}{3}& \frac23
\end{pmatrix}\\ 
   \mathbf{M}_3 &= \frac12\begin{pmatrix}
\frac13 &\frac{\sqrt2}{3} e^{-i2\pi/3} \\[0.2cm]
 \frac{\sqrt2}{3}e^{i2\pi/3}& \frac23\end{pmatrix}, \nonumber \\
 \mathbf{M}_4 &=\frac12\begin{pmatrix}
\frac13 &\frac{\sqrt2}{3} e^{-i4\pi/3} \\[0.2cm]
 \frac{\sqrt2}{3}e^{i4\pi/3}& \frac23
\end{pmatrix} \nonumber~.
\end{alignat}

We can also use measurements in the computational basis to construct  non-informationally complete representations of quantum states. This POVM leads to a EBM with an alphabet size of $q=2$. The POVM operators corresponding to computational basis measurements are given below,

\begin{align}
    \mathbf{M}_1 = ~\begin{pmatrix}
1 & 0 \\
0 & 0 
\end{pmatrix},~& \mathbf{M}_2 = ~\begin{pmatrix}
0 & 0 \\
0 & 1 
\end{pmatrix}.
\end{align}

For the experiment with GHZ states we use the tetrahedral POVM rotated by a Haar random unitary. This rotation is seen to improve performance in certain cases.

\section{Essential background on learning of EBMs}
\label{app:EBM_intro}
Like in any machine learning problem, learning the energy function becomes prohibitively hard if we don't restrict the hypothesis space of possible energy functions. Hence it is necessary to choose a parametric family of functions such that the learned model can be expressed as a member of this family, ideally using a small number of free parameters. In the case of representing quantum states, this parametric family can be chosen based on the prior information available about the state. In this study, we will work with polynomials with a fixed degree, feed-forward neural nets, and symmetric functions as parametric families for representing the energy functions.  We will demonstrate how each of these families is appropriate for learning representations for different classes of quantum states and POVMs.

Standard parametric estimation techniques like maximum likelihood are inefficient for learning the energy function, except within a very restricted function class \cite{chow1968approximating, mezard2009information}. Instead, we use a computationally efficient and sample optimal approach known as \textit{Interaction Screening} (IS) \cite{vuffray2016interaction, lokhov2018optimal, vuffray2019efficient, abhijith2020learning} for learning the energy function. For each spin in the model, this method finds a representation of the part of the energy function connected to that spin. For example, for the EBM  with an energy function $E(\ul{\sigma}) = \sum_{i,j} J_{ij} \sigma_i \sigma_j$, the IS method using a polynomial representation would learn the function $\sum_{j \neq i} J_{ij} \sigma_i \sigma_j$ for each $i$. We call this part of the energy function the \textit{local energy} associated with spin $\sigma_i.$

It is easy to see that for any EBM,  given the local energy for any $\sigma_i$ the conditional of this spin given every other spin $\mu(\sigma_i | \sigmab_{\setminus i})$ can be easily expressed, see \emph{SI}, section \ref{app:GM} for more details. These conditionals can then be used by a MCMC method to effectively generate samples from the learned EBM and to estimate any expectation value the POVM gives us access to \cite{mackay2003information,abhijith2020learning}. In this sense, the EBM methodology satisfies the two main conditions that any generative model for quantum states must satisfy: effective and practical algorithms for learning a representation and inferring quantities of interest from that learned representation. 

Now we will give a brief overview of the theory behind learning discrete EBMs. No generative model can learn every quantum state efficiently. For many neural net-based approaches there is no theoretical understanding of which states can be hard for a given method. The method we present here is different, as from extensive work on the IS method \cite{vuffray2016interaction, lokhov2018optimal, vuffray2019efficient} and from information-theoretic lower-bounds \cite{santhanam2012information} there is an exact characterization of which distributions are hard to learn. These results are best explained by considering the EBM defined in \eqref{eq:gibbs_dist} as a Gibbs distribution of a classical Hamiltonian on $n$ spins with $O(1)$ couplings and at an inverse temperature of $\beta$. It is known that the number of samples required to learn energy function in  the monomial basis scales as $e^{O(\beta d)} \log{n}$, where $d$ is the maximum number of interaction terms (monomials) that are connected to any one spin. This expression tells us that learning is easy for models at a fixed inverse temperature $\beta = O(1)$ defined locally on some lattice of fixed dimension $d = O(1)$.

\paragraph*{Selection of the parametric family.}

The EBM learning procedure outlined in the previous sections requires us to choose a parametric function family that can effectively capture the distribution generated by a quantum state/POVM combination. We find that this choice can be effectively made for a family of states by learning a polynomial representation for the energy function for a small state. The polynomial family can be chosen to include higher order terms which would be computationally infeasible to consider for larger states, as including all terms up to order $k$ would imply a computational cost of $O(n^k).$ Also, for smaller states it is possible to do learning in the limit of infinite number of samples drawn from the state( $m \rightarrow \infty$ limit) as the classical distribution generated by a POVM can be computed exactly for smaller states.  Learning the energy function in the polynomial basis for smaller states can give us valuable information about the types of terms present in the energy function and their symmetry properties. This information can then be used to choose the appropriate function family to use for larger states. If there are no obvious symmetries or if there are terms of higher order present, then a neural net family can be chosen to represent the energy function. If energy functions learned from smaller states only show the presence of terms of quadratic or lower order, then a polynomial ansatz would be suitable. If we observe translation or full permutation symmetry in the learned terms, then these symmetries can be incorporated to reduce the size of the parameter space we have to optimize over.

\section{Learning EBMs with Interaction Screening method}
\label{app:GM}

Given a set of $n$ random variables $\ul{\sigma} = \{ \sigma_1, \sigma_2, \ldots, \sigma_n  \},$ an EBM is a positive joint probability distribution over these variables given by,
\begin{equation}
    \mu(\ul{\sigma}) = \frac{\exp( E(\ul{\sigma}) )}{Z}. \end{equation} 
    
    Here $E$ is a real valued function known as the energy function  and $Z$ is the partition function that ensures normalization. The random variables can be continuous or discrete. In this work we will focus on the discrete EMBs. Moreover we assume that each random variable in the model takes values form the set $[q]$. We will  refer to $q$ as the alphabet size of the model.  For each variable, $\sigma_u$, in the model it is possible to split the energy function into two parts such that one of the parts completely capture the dependence of the energy function on the values taken by that variable alone,

    \begin{align} 
    E(\sigmab) &=  \langle \bm{\phi}(\sigma_u), \mb{E}^u(\sigmab_{\setminus u}) \rangle +  E^{\setminus u } (\sigmab_{\setminus u}) \label{eq:local_en_decomp} \\ 
    &=  \sum^q_{a=1}\phi_a(\sigma_u)( ~E^u_a(\sigmab_{\setminus u}; \thetab) )+    E^{\setminus u } (\sigmab_{\setminus u})
\end{align}

Here, $\phi_a(x) =  \delta_{a,x} -  1/q$ are centered delta functions, and  $\bm{E}^u: [q]^{n-1} \rightarrow \mathbb{R}^q$ is a vector valued function. Together  $\langle \bm{\phi}(\sigma_u), \mb{E}^u(\sigmab_{\setminus u}) \rangle$ includes in it all the terms of $E$ that are dependent on $u$. We call this term the \textit{local energy function of $u$}.   The second term, $E^{\setminus u}(\sigmab_{\setminus u})$  includes the rest of the terms  that don't depend on the value taken by $\sigma_u.$ Notice that this decomposition does not put any restriction on our EBM as  any multivariate function with discrete inputs can be decomposed in this way. 

The local energy of a variable $\sigma_u$ is related to its conditional distribution given all the other variables:
\begin{equation}\label{eq:conditional}
\mu[ \sigma_u | \ul{\sigma}_{\setminus u} ] = \frac{  \exp (\langle \bm{\phi}(\sigma_u), \mb{E}^u(\sigmab_{\setminus u}) \rangle)  }{\sum_{s=1}^q \text{exp} ( \langle \bm{\phi}(s), \mb{E}^u(\sigmab_{\setminus u}) \rangle )    }. 
\end{equation}

This makes it possible to sample from the EBM using Gibbs sampling once local energies of every variables are known.

To learn a EBM simply means to learn a representation for its energy function. We achieve this using the interaction screening method, first introduced in \cite{vuffray2016interaction}. This method can be used to find representations for the local energies of each variable in the model. Given $m$ i.i.d samples $\{\sigmab^{(1)},\ldots, \sigmab^{(m)}\}$ drawn from a EBM, these representations are learned by minimizing the following loss function for each variable in the model,

\begin{align}
\label{eq:objective}
\s_u(\thetab) &= \frac{1}{m} \sum_{t = 1}^m \exp ( -\langle \bm{\phi}(\sigma^{(t)}_u), \mb{f}(\sigmab^{(t)}_{\setminus u}; \thetab) \rangle  ) \\
&=  \frac{1}{m} \sum_{t = 1}^m \exp (- \sum^q_{a=1}\phi_a(\sigma^{(t)}_u) ~f_a(\sigmab^{(t)}_{\setminus u}; \thetab)  ).
\end{align}

Here, $\bm{f}(; \thetab): [q]^{n-1} \rightarrow \mathbb{R}^q$ is a vector valued function parametrized by $\thetab$.  For example, $\bm{f}(; \thetab)$ can be an affine transformation from $\mathbb{R}^{n-1}$ to $\mathbb{R}^q$. In this case $\thetab$ would be the set of coefficients that define the transformation.

The value of $m$ required in \eqref{eq:objective} to learn a polynomial energy function, up to a certain error, can be bounded using techniques from stochastic convex optimization. The complete analysis can be found in \cite{vuffray2019efficient} and the sample complexity of this method is found to asymptotically match known lower bounds \cite{santhanam2012information}.

We can get an intuition on how the IS method works by taking $m \rightarrow \infty$ limit in \eqref{eq:objective}.  Also assume that for every variable, $\sigma_u$, in the EBM there exists some $\thetab^{u*}$ such that $\mb{f}(;\thetab^{u*}) = \mb{E}^u$. This condtition implies that the parametric function family is expressive enough to capture all the local energies of the true EBM. Under these assumptions one can easily see that the gradient of $\s_u(\thetab)$ with respect to the $\theta$ variables vanish at $\thetab^{u*}.$ Moreover, one can show that these stationary points are the global minima of $\s_u(\thetab)$ \cite{abhijith2020learning}. 

The above argument motivates the use of the interaction screening loss function for learning EBMs. Even for a finite number of samples, we expect the global minima of $\s_u(\thetab)$ to encode useful information about the local energies of the EBM. Now using this learned representation for the local energies we can compute the single variable conditionals of the learned EBM using Eq.\eqref{eq:conditional}. Once the single variable conditionals are known we can sample from the learned model using Gibbs sampling.

The versatility of the interaction screening method comes from the fact that any parametric function family can be chosen in the definition of the interaction screening loss. Then  by minimizing this loss we will be able to find a representation for the EBM in this function family. This freedom in choosing the function family lets us learn parsimonious representations for the EBM and incorporate prior information into the learning process.

\section{Parametric function families}
\label{app:FFamily}

Here, we discuss the parametric function families that we use for modeling the EBM energy functions in this study.

\subsection*{Polynomials}

Polynomials are the most widely used function family for representing EBMs. For discrete random variables, any energy function can be expressed as a linear combination of a finite number of monomial terms. For the monomial family, the minimization of the interaction screening objective is a convex optimization problem \cite{vuffray2016interaction, vuffray2019efficient, lokhov2018optimal}. The interaction screening method is also known to be sample optimal in this case. We train the polynomial models using Entropic Descent algorithm \cite{vuffray2019efficient, beck2003mirror}.

If the true energy function is  a polynomial function of degree $L$, then the complexity of learning a polynomial representation scales as $O(n^L)$. Thus learning polynomial representations are tractable only if the true energy  is given by a low-degree polynomial. Many models used in classical statistical physics have a low degree and can be efficiently learned using polynomials. But this low degree condition is not satisfied in general for quantum tomography data. 

\subsection*{Neural nets}

The use of neural nets to in the interaction screening method was introduced in \cite{abhijith2020learning}. The neural network anzatz is useful if the EBM being learned has higher order terms and if there are no obvious symmetries in the problem  that can reduce the size of the polynomial ansatz. The interaction screening loss function for this case simply reads,

\begin{equation}
\label{eq:objective_nn}
\s_u(\thetab) = \frac{1}{m} \sum_{t = 1}^N \exp ( -\langle \bm{\phi}(\sigma^{(t)}_u), \mb{NN}(\sigmab^{(t)}_{\setminus u}; \thetab) \rangle  ). 
\end{equation}

Here we have used a vector valued neural net for the general function in Eq.\eqref{eq:objective}. The $\thetab$ parameters are now the weights and biases of the neural net. Since neural nets can approximate any arbitrary function one can show that this ansatz is powerful enough to learn any energy function. 

The minimization of this loss function can be done using a variant of the  stochastic gradient descent algorithm. The gradients can be computed using backpropagation. This enables the training of the nerual network ansatz on a GPU.

We use feed-forward neural nets as our neural net anstaz. In general, such a net is  a function, $f:\RR^{n-1} \rightarrow \RR^q$, defined as,
\begin{equation}
    \mathbf{f} (\xb) = \Ab_{d+1}\cdot \mathbf{\alpha} \left( \ldots \left( \Ab_2\cdot \left( \mathbf{\alpha} \left( \Ab_1\xb + \ul{b}_1  \right) \right) + \ul{b}_2 \right)\ldots \right)  + \ul{b}_{d+1}.
\end{equation}

Here $\Ab\cdot\xb + \ul{b}$ is an affine transformation and $\alpha: \RR \rightarrow \RR $, known as the activation function, acts on a vector component wise, i.e $ \mathbf{\alpha}(\xb) \equiv [\alpha(x_1), \alpha(x_2), \ldots ]$. The matrices $\Ab_1, \ldots, \Ab_{d+1},$ and the vectors $\ul{b}_1, \ldots, \ul{b}_{d+1}$, are the trainable parameters of this model.

We call the number $d$, that fixes the number of layers of the net, the \emph{depth} of the network. To simplify the architecture of the nets, we will force the intermediate layers to all have the same size, i.e $\Ab_1 \in  \RR ^{n-1\times w}$, $\Ab_2 \ldots \Ab_{d} \in \RR ^{w\times w}$, and $\Ab_{d+1} \in \RR^{w \times d}.$ We call $w$, the \emph{width} of the net.

Through out this work we will use the swish function as our activation function \cite{ramachandran2017swish}, $\alpha(x) =  \frac{x}{1 + \exp(-x)}.$ This is a differentiable variant of the popular ReLU activation function used extensively in deep learning. We will train our model using a popular variant of the Stochastic Gradient Descent algorithm known as ADAM \cite{kingma2014adam}. 

\subsection*{Symmetric functions}

Symmetric functions can be used as ansatze for the local energies to learn a fully symmetric EBM. If we have prior information that the data we use for learning comes from a fully symmetric distribution, then using such an ansatz can reduce the size of the model being learned from $O(q^n) $ to $O(qn^{q-1})$. This reduction in size makes the learning problem tractable.

A fully symmetric graphical model will be invariant under any permutation of its variables.
\begin{equation}
    \mu(\sigma_1 , \sigma_2,  \ldots, \sigma_n) =  \mu(\sigma_{\pi(1)} , \sigma_{\pi(2)},  \ldots, \sigma_{\pi(p)}), ~~\forall ~ \pi \in S_n.
\end{equation}

Here $S_n$ is the symmetric group on $n$ elements.

It is straightforward to see that all the single variable conditionals of this distribution will inherit this  permutation symmetry,
\begin{equation}
    \mu(\sigma_u | \ul{\sigma}_{\setminus u}) =  \mu(\sigma_u | \ul{\sigma}_{\pi(\setminus u)}), ~~\forall ~ \pi \in S_{n-1}.
\end{equation}

This implies that the true local energies of the model can be represented by symmetric functions on $n-1$ variables and we can use this function family in the interaction screening loss function. The outputs of such functions are agnostic to the order of their inputs, i.e they depend only the number of occurrences of each alphabet. This property can be used to find compact representations for such functions. Let us define a function that counts the occurrence of each alphabet in an input, $\sharp:[q]^{n-1} \rightarrow \mathbb{R}^q$, such that $\sharp(\ul{\sigma}_{\setminus u})_a$ gives the number of occurrences of $a$ in $\ul{\sigma}_{\setminus u}$. Using this we can define the following interaction screening loss function ,

\beq\label{eq:symmetric_iso}
\s(\mb{\Theta}) = \frac{1}{N} \sum_{t = 1}^N \exp ( -\langle \bm{\phi}(\sigma^{(t)}_1), \mb{\Theta} ( \sharp(\sigmab^{(t)}_{\setminus 1})) \rangle  ),
\eeq

Here $\mb{\Theta}$ is a vector-valued function that acts as our local energy ansatz. The possible inputs to $\mb{\Theta}$ are integer strings of length $q$ that sum to $n-1$. There are only $O(n^{q-1})$ such inputs and we can use the outputs of $\mb{\Theta}$ at each possible input as the optimization variables in our loss function. Also using these optimization variables leads to a convex loss function and hence fast optimization in practice. The optimization need only be done for the first spin as all the local energies will be the same due to the symmetry of the model. We perform this optimization using interior-point methods provided in the Ipopt package \cite{wachter2006implementation}. This is guaranteed to find the global optimum of the IS loss function.

Other symmetries can also be incorporated in a similar way by using a variational ansatz for the energy  function that is invariant under the symmetry transformations in question. For the any parametric function family, if the symmetry group is small enough, this can be achieved by simple averaging. For instance, a new function $\tilde{\mathbf{f}}$, defined as,
$\tilde{\mathbf{f}}(\ul{\sigma}, \theta) \equiv \mathbf{f}(\ul{\sigma}, \theta)  + \mathbf{f}( - \ul{\sigma}, \theta),$ is by construction symmetric under spin flip and is useful for learning models with a global $\mathbb{Z}_2$ symmetry. 

It is also worth noting that the interaction screening objective can naturally impose translation invariance with any function family. This is because optimizing the objective  \eqref{eq:objective} for any  function family reconstructs the local piece of the energy function at each site. If the model is transitionally invariant, every local piece of the energy function will be the same (i.e.  $\mathbf{E}^u =  \mathbf{E}^v ~ \forall ~ u,v \in [n] $) and we only need to optimize for a single site.

\section{Additional details about methodology}
\label{app:Methodology}

\subsection*{Additional details for numerical experiments}

In this section, we provide details on the experiments involving learning of EBM representations of quantum systems.  

\begin{enumerate}
    \item[\figurename \ref{fig:order_gibbs_POVM}] Exact distribution $\mu$ was obtained by exactly exponentiating the quantum Hamiltonian. This distribution was used to take the $m \rightarrow  \infty$ limit in the IS loss function. In the learning process, a third-order polynomial ansatz was optimized using an interior point method from Ipopt.

    \item[\figurename \ref{fig:Corr_1D_TIM_NN}] This experiment uses a neural nets with depth, $d=3$ and width, $w =15$. The model was trained with a mini-batch size of $500$, with an early stopping criteria with a delta of $10^{-4}$ for a maximum of $1500$ epochs. Training was done using ADAM with initial learning rate of $10^{-2}$ and an inverse time decay schedule.

    \item[\figurename \ref{fig:Corr_1D_TIM_ground}, \figurename \ref{fig:gs_scaling}] This experiments in these figures uses a neural nets with depth, $d=3$ and width, $w =25$. The model was trained with a mini-batch size of $5000$, with an early stopping criteria with a delta of $10^{-8}$ for a maximum of $500$ epochs. Training was done using ADAM with initial learning rate of $8\times 10^{-3}$ and an inverse time decay schedule.
    
    \item[\figurename \ref{fig:order_gibbs}:] Exact distribution $\mu$ was obtained by exactly exponentiating the quantum Hamiltonian. In the learning process, a polynomial ansatz with terms at all orders was optimized using Entropic GD.

    \item[\figurename \ref{fig:TV_GHZ_compare}]
    This experiment uses a neural nets with depth, $d=2$ and width, $w =8$. The model was trained with a mini-batch size of $3000$, for $200$ epochs. Training was done using ADAM with initial learning rate of $10^{-2}$ and an inverse time decay schedule.
    
    \item[\figurename \ref{fig:order_ground}:] Exact distribution $\mu$ was computed from an MPS representation of the ground state obtained by DMRG. In the learning process, a third-order polynomial ansatz was optimized using an interior point method from Ipopt.
    
    \item[\figurename \ref{fig:2D_corr_Heisenberg}] This experiments in these figures uses a neural nets with depth, $d=3$ and width, $w =15$. The model was trained with a mini-batch size of $10000$, with an early stopping criteria with a delta of $10^{-8}$ for a maximum of $500$ epochs. Training was done using ADAM with initial learning rate of $8\times 10^{-3}$ and an inverse time decay schedule.
    
\end{enumerate}

\subsection*{Error metrics}

The interaction screening method is nearly sample-optimal for learning the parameter values in the energy function \cite {lokhov2018optimal}. But for data generated from quantum states, the true parameter values of the energy function are not known. Extracting the parameters in the energy function can be done by brute force for smaller systems, but becomes prohibitively expensive even for systems of moderate size. Thus to compare the error in our learned representations we compute the error in one and two body observables in $L_1$ norm, averaged over the entire lattice. For models learned in  the computational basis, this reduces to average errors in absolute value in observables of the form $\langle Z \rangle$ and  $\langle ZZ \rangle$. For models learned using samples generated by the tetrahedral POVM, we use the trace distance between one and two body reduced density matrices as our error metric. This quantity then upper bounds the average error in any one and two body observable, up to a constant factor \cite{nielsen2002quantum}.

\section{Estimates for quantum fidelity}
\label{app:Fidelity}

Given a pure state $\ket{\psi}$ and a mixed state $\rho$, we define the quantum fidelity between these states as the following overlap \cite{jozsa1994fidelity},
\begin{equation}
    F^Q({\ket{\psi}},\rho) \equiv \braket{\psi|\rho|\psi}.
\end{equation}

Expanding $\rho$ in the dual POVM elements as in \eqref{eq:dual}, we get,
\begin{equation}
    F^Q({\ket{\psi}},\rho) = \sum_{\ul{\sigma}} \mu(\sigmab) \braket{\psi|D_{\sigmab}|\psi}
\end{equation}

This lets us construct a finite sample estimate for $F^Q$ given $N$  samples from the distribution $P$  just as in \eqref{eq:approx},
\begin{equation}
    F^Q_N ({\ket{\psi}},\rho)  =  \frac{1}{N} \sum_{k=1}^N \braket{\psi|  ~ \mb{D}^{(1)}_{\tau^{(k)}_1} \otimes \mb{D}^{(2)}_{\tau^{(k)}_2} \otimes \ldots \otimes \mb{D}^{(n)}_{\tau^{(k)}_n} | \psi} 
\end{equation}

This gives an unbiased estimate for the quantum fidelity  and can be easily estimated in cases where $\psi$ has an MPS representation or when it can be written down as a superposition of a few computational basis states.

\section{Additional numerical results}
\label{app:Additional_numerics}

\subsection*{Scaling experiments for thermal states}

\begin{figure*}[h]
 \vspace{-0.1in}
\centering
\begin{subfigure}[b]{0.35\textwidth}
  \caption{1D TIM, $\beta = 1.0, g = 1.0$}
    \includegraphics[width=\textwidth]{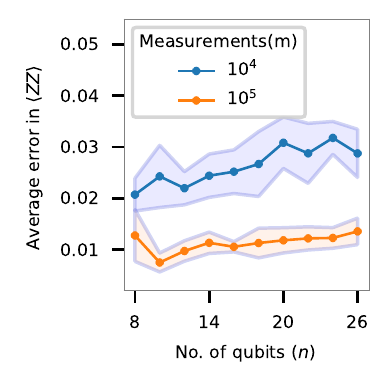}
    \label{fig:scaling_1D_a}
\end{subfigure}    
\begin{subfigure}[b]{0.35\textwidth}
  \caption{1D TIM, $\beta = 1.0, g = 1.0$}
    \centering
    \includegraphics[width=\textwidth]{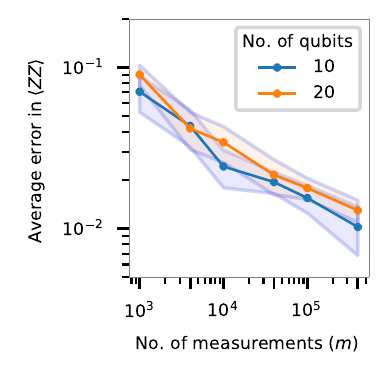}
    \label{fig:scaling_1D_b}
\end{subfigure}  \\
  \begin{subfigure}[b]{0.35\textwidth}
  \caption{2D TIM, $\beta = 1.0, g = 1.0$}
    \includegraphics[width=\textwidth]{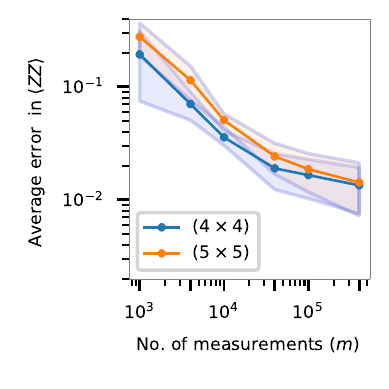}
    \label{fig:samples_vs_err_gibbs_2D}
\end{subfigure}    
\begin{subfigure}[b]{0.35\textwidth}
\caption{2D TIM, $\beta = 1.0, g = 3.5$}
    \includegraphics[width=\textwidth]{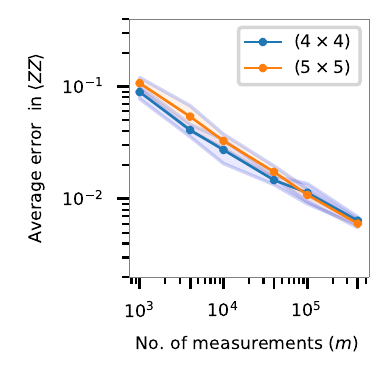}
    \label{fig:samples_vs_err_gibbs_2D_critical}
\end{subfigure}
\caption{{\bf Scaling results for the classical representations of thermal states of TIM.} Error in ZZ expectation of the polynomial representation for the 1D TIM at with  $\beta=1$  {\bf (a)} number of qubits {\bf (b)} number of measurements (i.e. training samples). We see that increasing the number of measurements has decreased the error significantly.  On the other hand, the error does not increase appreciably if the number of qubits in the system is increased. {\bf (c)} Error in ZZ expectation of the polynomial representation for the 2D TIM at $\beta=1$ with  $g = 1$. {\bf (d)} The same but for $g = 3.5 $. The chosen $g$ values are on either side of the zero-temperature critical point ($g_c = 3.044)$ \cite{sachdev_2011}. We see that the model is easier to learn for the larger value of $g$.}
\label{fig:scaling_gibbs_2D}
\end{figure*}

Scaling results for learning thermal states using the polynomial ansatz are given in \figurename~\ref{fig:scaling_gibbs_2D}. \figurename~\ref{fig:scaling_1D_a} and \figurename~\ref{fig:scaling_1D_b} show results for a 1D TIM.  We can see that error in the expectation values is most significantly affected by the number of samples used for learning. On the other hand, increasing the number of qubits in the system does not increase the average error in the expectation values by much. \figurename~\ref{fig:samples_vs_err_gibbs_2D} and \figurename ~\ref{fig:samples_vs_err_gibbs_2D_critical} show similar scaling experiments for a 2D TIM.  Here we see that error has only a weak dependence on the size of the system. The error is seen to decrease consistently with the number of training samples. But the model with a higher value of $g$ is seen to be easier to learn. This points to an effective temperature in the EBM induced by the transverse field. 

\subsection*{Additional experiments with ground states}

\begin{figure*}[h]
    \centering
     \begin{subfigure}[t]{0.35\textwidth}
    \caption{}
    \includegraphics[width=\textwidth]{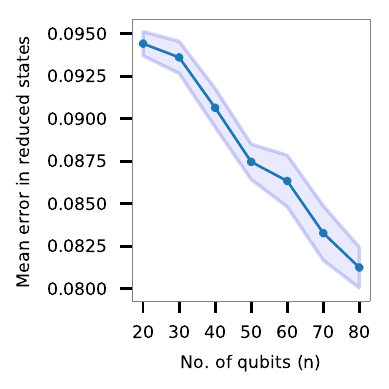}
    \label{fig:my_label}
\end{subfigure}    
     \begin{subfigure}[t]{0.35\textwidth}
    \caption{}
    \includegraphics[width=\textwidth]{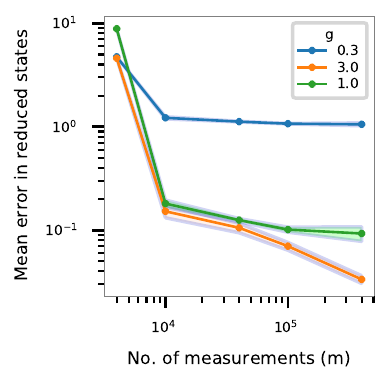}
\label{fig:err_vs_g}
\end{subfigure}    
    \caption{{\bf Scaling results for the classical representations of ground states of TIM.} {\bf (a)} Error in one and two body reduced states as a function of number of qubits for ground state of 1D TIM. The average error goes down in this case  as most errors occur when the spins are close together in the lattice. These EBMs are learned using $m=10^5$ samples. {\bf (b)} For a $60$-qubit 1D TIM model, we see the effect of $g$ on learning. The low-$g$ state here is hard to learn as it is close to the ground state of a classical model. This implies that the effective temperature of the EBM is low and the sample complexity of learning is high \cite{santhanam2012information}. On the other hand  higher $g$ states are easier to learn, as evidenced by the decreasing error as the number of training samples is increased. Presented is the error (average trace distance between one- and two-qubit reduced states) as a function of the number of training samples $m$.}
    \label{fig:gs_scaling}
\end{figure*}

Scaling experiments for learning ground states is given in \figurename~\ref{fig:gs_scaling}. Interestingly in \figurename~\ref{fig:my_label}, we see that the average error in reduced states decreases with increasing system size. This is due to fact that most errors for two-qubit reduced states occur when the qubits are closer together on a lattice. For qubits that are farther apart, the EBM reconstructs the reduced states well.

In \figurename~\ref{fig:err_vs_g} we study the effect of the phase of the ground state on the learning algorithm for a 1D TIM. Below the critical point of $g = 1$, the ground state is close to the ground state of the classical Ising model. As explained in \textit{Supplementary materials} , these states are known to be hard to learn. We see that reflected in these experiments, as the errors in the reduced do not decrease when we increase the training samples. For higher values of $g$, the ground state moves away from the classical ground state and this allows for easier learning. This is evidenced by a consistent decrease in error with increasing training samples in these models.

\begin{figure*}[h]
    \centering
    \includegraphics[width=0.65\textwidth]{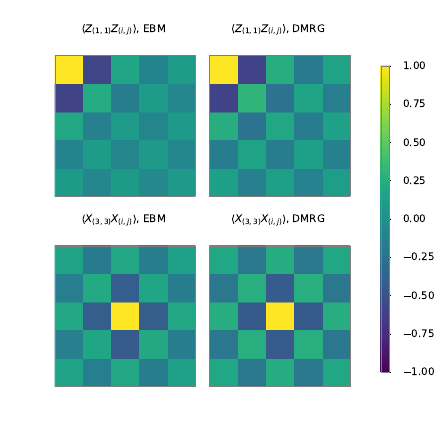}
    \caption{{\bf Learning classical EBM representation for the ground state of a 2D Heisenberg model.} Comparison of $ZZ$ and $XX$ expectation values learned by the NeuRISE algorithm with values obtained from DMRG, $m=4 \times 10^6$. The Hamiltonian here is the anti-ferromagnetic Heisenberg model on a $5\times 5$ lattice. We see good agreement in local observables for learning this 2D model. And just like for the 1D models discussed in the main text, we also see improved predictions while increasing the training values of $m.$ The neural net used here has only $23250$ free parameters compared to the Hilbert space dimension of $ \approx 3 \times 10^7$.}
    \label{fig:2D_corr_Heisenberg}
\end{figure*}

We test the quality of the EBM representation for ground state of a model other than TIM in \figurename~\ref{fig:2D_corr_Heisenberg}, where we study the Heisenberg model in 2D.
\begin{equation}
    H = \sum_{\langle  i,j \rangle} X_iX_j + Y_iY_j + Z_i Z_j.
\end{equation}

Here we use a neural net representation for the energy function. The correlation functions obtained from the learned EBM representation show an excellent agreement with the ground-truth correlation expectation values.

\end{document}